\documentclass{aa}  
\usepackage{graphicx}
\usepackage{subfigure}
\usepackage{txfonts}
\usepackage{hyperref}
\usepackage[final]{changes}  

\begin{document}
 \title{Multiwavelength spectroscopic observations of a quiescent prominence }
   \author{Jianchao Xue\inst{1}\fnmsep\thanks{The first two authors are listed in alphabetical order but have equally contributed to the paper}
          \and
          Ping Zhang\inst{1}\fnmsep\footnotemark[1]
          \and
          Jean-Claude Vial\inst{2}\fnmsep\thanks{Corresponding author: jean-claude.vial@universite-paris-saclay.fr, lfeng@pmo.ac.cn}
          \and
          Li Feng\inst{1}\fnmsep\footnotemark[1]\footnotemark[1]
            \and
          Maciej Zapi{\'o}r\inst{3}
              \and
           Werner Curdt \inst{4}
              \and
           Hui Li \inst{1}
              \and
           Weiqun Gan \inst{1}
          }

   \institute{Key Laboratory of Dark Matter and Space Astronomy, Purple Mountain Observatory, Chinese Academy of Sciences, Nanjing 210023, China
         \and
             Institut d'Astrophysique Spatiale, CNRS/Université Paris-Sud, Université Paris-Saclay, 91405, Orsay, France
           \and
            Astronomical Institute of the CAS, Fričova 298, 25165 Ondřejov, Czech Republic
            \and
             Max Planck Institute for Solar System Research, 37191 Katlenburg-Lindau, Germany\\
             }
  
   \date{Received ; accepted }
   \titlerunning{Multiwavelength spectroscopic observations of a quiescent prominence}
    \authorrunning{Xue et al.}

   \date{Received xxx; accepted xxx}
   \abstract
   {}
    {In this paper we focus on the analysis of the multiwavelength spectroscopic observations of a quiescent prominence. The spectral and geometrical parameters in the prominence were derived and used to constrain the nonlocal thermodynamic equilibrium (NLTE) radiative transfer models of the prominence. Applying this method with multiwavelength observations provides a good opportunity to reduce the large range of thermodynamic parameters in solar prominences.}
   {As far as velocities are concerned, we used time-slice and optical flow methods in order to derive the plane-of-sky (POS) velocities in the prominence, and used gravity center and peak position methods on \ion{Mg}{ii} h\&k and \ion{H}{i} Ly$\alpha$ profiles to compute the line-of-sight (LOS) velocities. As far as densities and temperatures are concerned, we used the integrated intensities and full width at half maximum (FWHM) values of the H$\alpha$ and the \ion{Ca}{ii} H together with \ion{Mg}{ii}\ h\&k lines to compare with values derived from the NLTE radiative transfer computations. Ionization degree and thickness of the prominence plasma could be further derived.}
   {Opposite flows are observed along two strands between prominence barbs. The POS velocity can reach $20\,\mathrm{km\,s^{-1}}$ and the largest LOS velocity is $>90\,\mathrm{km\,s^{-1}}$.  The derived electron densities range from $6.5\times10^9$\,cm$^{-3}$ to $2.7\times10^{10}$\,cm$^{-3}$ and the derived total hydrogen densities range from $7.4\times10^9$\,cm$^{-3}$ to $6.6\times10^{10}$\,cm$^{-3}$ in different regions of the studied prominence. The temperature ranges from $7\,000\,\mathrm{K}$ to 14\,000\,K. The ionization degree of hydrogen is in the range of 0.40 to 0.91. The comparison between averaged and modeled profiles of \ion{Mg}{ii} and Ly$\alpha$ lines shows that macro-velocities of $15\,\mathrm{km\,s^{-1}}$ and $20\,\mathrm{km\,s^{-1}}$ are required, respectively.} 
   {The bulk motions among prominence barbs indicate that the prominence plasma is not confined within magnetic dips but exhibits a large-scale behavior. The presence of high-speed cool plasma flows, along with a wide range of plasma densities and temperatures, suggests that the prominence plasma is far from thermodynamic equilibrium and is inherently dynamic in nature.}
   \maketitle
\makeatletter
\ifaa@referee
    \setlength{\highfigwidth}{0.8\linewidth}
\else
\fi
\makeatother  
  
\section{Introduction}
Solar prominences are very large-scale cool and dense plasma structures supported and confined in the hot solar corona by magnetic fields \citep{tandberg1995astrophysics, mackay10a,vial15book}. The prominence plasma is two orders of magnitude cooler and denser than its coronal surroundings \citep{labrosse10b}. The temperature inside the prominence is not uniform, the interface region between the hot corona and the cool core of the prominence is usually called the Prominence Corona Transition Region \citep[PCTR, ][]{vial1990prominence}. Measuring the physical parameters of solar prominences such as the kinetic temperature, effective thickness, electron density, and microturbulent velocities is essential for the study of  the  physical  state  of  the  prominence  and  also  for  the development and testing of models describing the prominence internal  structure. 

The prominence emission is observable in a broad spectral range and consists of various lines formed at chromospheric to coronal temperatures, such as the neutral hydrogen H$\alpha$ and Ly$\alpha$ lines and \ion{Mg}{ii}\ h\&k lines. With the advent of continuous high-resolution observations and spectroscopic imaging measurements from satellites (e.g., the Solar Ultraviolet Measurements of Emitted Radiation  \citep[SUMER;][]{wilhelm1995sumer} on board the Solar and Heliospheric Observatory (SOHO) , the Atmospheric Imaging Assembly \citep[AIA;][]{lemen12}  telescope on board the Solar Dynamics Observatory \citep[SDO;][]{pesnell12} and the Interface Region Imaging Spectrograph \citep[IRIS;][]{depontieu14}) and ground-based solar telescopes (e.g., the Solar Horizontal Spectrograph \citep[HSFA2;][]{kotrc2007modernized} of Ondřejov Observatory), much progress has been made in the last decades in statistical studies of the prominences and helps to understand their general properties. In the last decade, more than 20 papers involving IRIS observations have been devoted to solar prominences as early as 2014 \citep{2014AA...569A..85S,vial16}. They concern various structures: mostly quiescent, but also erupting \citep{2015ApJ...803...85L,2019AA...624A..72Z,2021FrP.....9..543X} and  horn-like \citep{2023AA...680A..63B}. These studies concern velocity fields \citep{2014AA...569A..85S,2016ApJ...831..126O,2017AA...606A..30S,2018ApJ...865..123R} and/or diagnostic of the plasma through NLTE (departure from the local thermodynamic equilibrium) modeling \citep{2015SoPh..290..381V,2021A&A...653A...5P,2022ApJ...932....3J,2023A&A...679A.156P,2023AA...680A..63B}. \citet{2018A&A...618A..88J} addressed the relationships between intensities and optical thicknesses of \ion{Mg}{ii}, \ion{C}{ii} and \ion{Si}{iv} lines of a quiescent prominence with IRIS observations, and the results suggested that the intensity variations are caused by random motions rather than global oscillations, and the \ion{Mg}{ii} and \ion{C}{ii} lines from the prominence are generally optically thick. \citet{2022ApJ...934..133G} emphasized the significant impact of incident radiation and dynamic motions on the radiative transfer modeling of \ion{Mg}{ii} k and h lines emergent from prominence-like structures.

Given that spectral widths are mainly contributed from thermal and nonthermal motions along the line of sight (LOS), spectral pairs from different atoms can be used to constrain plasma temperatures and microturbulent velocities. Utilizing ground-based spectral observations, \citet{2020PASJ...72...71O} fitted spectral profiles with a cloud model, and found that the pairs of \ion{H}{i} H$\beta$ and \ion{Ca}{ii} $8542\,\mathrm{\AA}$ lines could constrain plasma temperature accurately. The derived kinetic temperature of solar prominences ranges from $8\,000\,\mathrm{K}$ to $12\,000\,\mathrm{K}$. However, ignoring the effect of large opacity on the spectral broadening would overestimate the derived temperature and microturbulent velocity. Because of the low temperature and high density, the plasma in the cool core of the prominence is partially ionized, thus it is optically thick in most of the resonance lines and continua of hydrogen and helium \citep{labrosse2004non}. One has to solve the sophisticated NLTE radiative-transfer problem to derive the thermodynamic parameters from the observed line and continuum intensities. 

Quiescent prominences are usually modeled as one-dimensional (1D) vertical plasma slabs illuminated by the surrounding atmosphere. Such modeling can be applied to investigate the radiative properties of isothermal-isobaric slabs with varying pressures, temperatures, and thicknesses, and can also be used to derive relations between model parameters and the emitted lines and continua \citep{1976HeasleyMihalas, heinzel1987formation, 1993GHV}. In addition to modeling H spectra, this method has been extended to include atoms such as \ion{He}{i}, \ion{He}{ii}, \ion{Ca}{ii}, and \ion{Mg}{ii} \citep{1993Paletou, labrosse2012plasma}. Although the fine structure of prominences is well supported by previous high-resolution observations \citep{lin2005thin}, a 1D slab remains a realistic approximation in prominence modeling, given that the emitted spectrum is primarily affected by the central cool regions. 

By comparing observed profiles with those derived from NLTE radiative transfer calculations, a great number of plasma parameters can be constrained. Using observations and modeling of the  \ion{Mg}{ii} h\& k lines, \citet{2019AA...624A..72Z} found that the mean temperature of an erupting prominence is around $11\,000\,\mathrm{K}$, and the electron density is in the range of $1.3\times10^9$ to $6.0\times 10^{10}\,\mathrm{cm^{-3}}$. For a quiescent prominence, \citet{2022ApJ...932....3J} found that the temperature is below $10\,000\,\mathrm{K}$, the electron density is on the order of $10^{10}\,\mathrm{cm^{-3}}$, and the microturbulent velocity is typically $8\,\mathrm{km\,s^{-1}}$. For simplification, 1D single-slab models are commonly used to constrain the observations. However, we should note that 2D (e.g., a more penetrating incident radiation) and multi-slab (e.g., different motions along the LOS) effects are ignored in those works. For a presentation of the pros and cons of the various modelings, we refer the reader to \citet{Heinzel2025radiative}. In addition to spectral observations, \citet{2022ApJ...927L..29H} found that ALMA, with observations at the wavelength of $3\,\mathrm{mm}$, is a prominence thermometer for bright parts. In spite of those measurements, there is still a large uncertainty in the range of thermodynamic parameters of the prominence plasma. 

In the present paper, we study the physical conditions of a quiescent prominence observed on March 28, 2017 at the northeast limb with a set of observations including IRIS, SUMER, and HSFA2. We introduce the observations and the corresponding instruments in Section~\ref{sec:obs}. Section~\ref{sec:vel} shows our calculations and analysis of the velocity pattern in the prominence. Section~\ref{sec:dia} details the setup of the NLTE models and the diagnostics of the prominence plasma. Section~\ref{sec:dis} summarizes our conclusions from this work.

\section{Observations} \label{sec:obs}

\begin{figure*}[hbt]
  \centering
   \includegraphics[width=.75\linewidth]{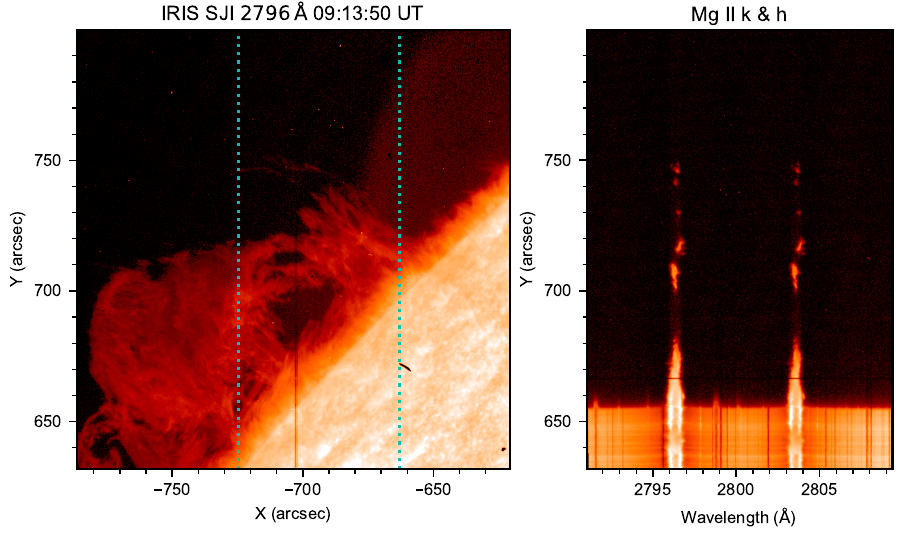}\hfill
   \caption{Example of IRIS observations. Left: IRIS SJI in $2796\,\mathrm{\AA}$. The dark line in the center corresponds to the position of the spectrograph slit, and the dotted cyan lines mark the two extreme positions of the slit. Right: \ion{Mg}{ii}\ h\&k spectra taken along the slit in (a). Both the two images are shown in logarithmic scale.}
   \label{fig:irisiandp}  
\end{figure*}
  
\begin{figure*}[hbt]
 \centering
  \includegraphics[width=.8\linewidth]{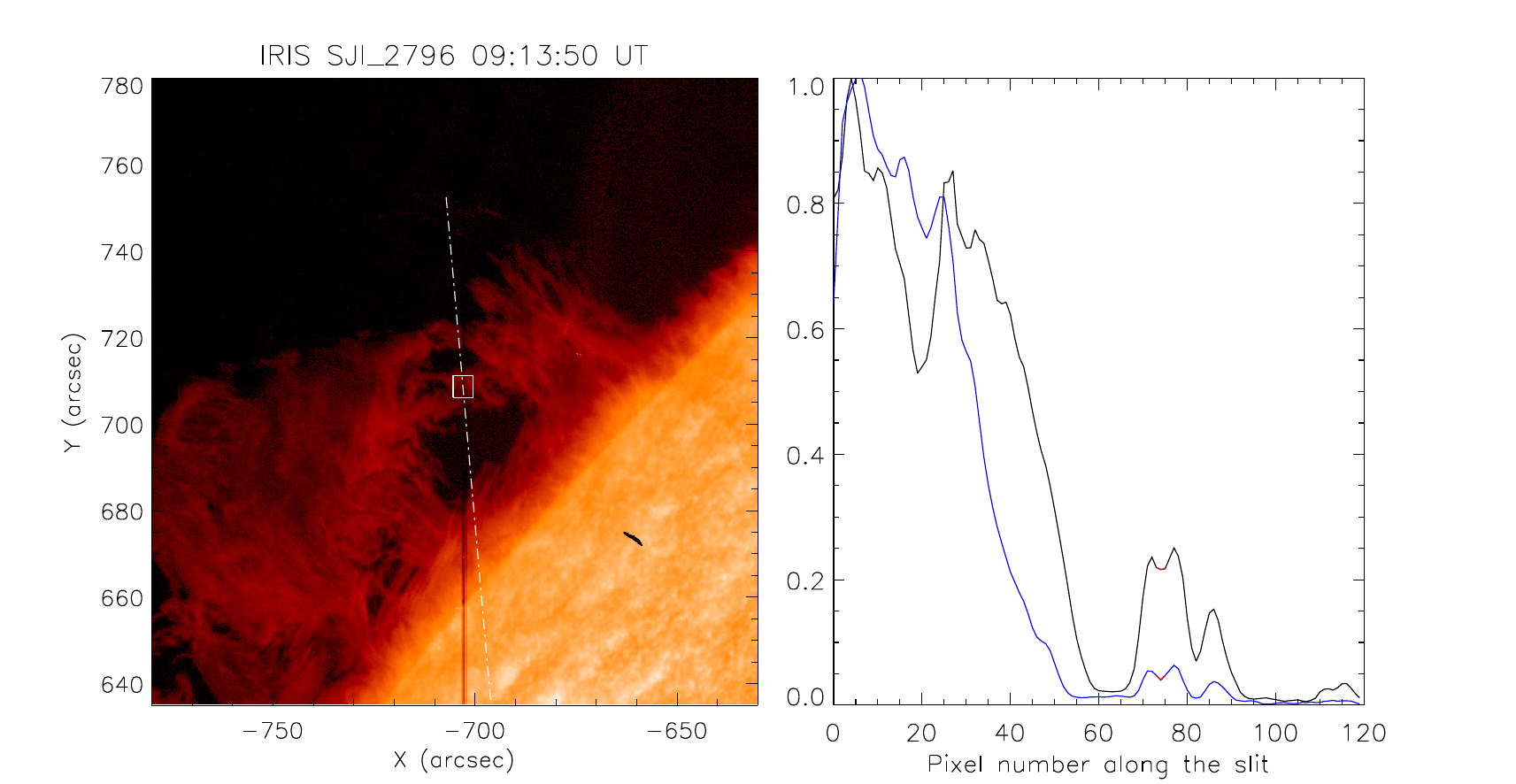}\hfill
  \caption{Coalignment between the IRIS and SUMER observations. The dash-dotted line in the left panel in white represents the position of the SUMER slit. The right panel shows how the normalized integrated intensity changes along the slit. The blue line corresponds to the intensity of the pixels of the IRIS SJI along the white line from the bottom to the top. The black line corresponds to the Ly$\alpha$ integrated intensity at 09:13:33 UT along the SUMER slit.}
  \label{fig:irisandsumer}  
\end{figure*}

\begin{figure*}[hbt]
 \centering
  \includegraphics[width=.8\linewidth]{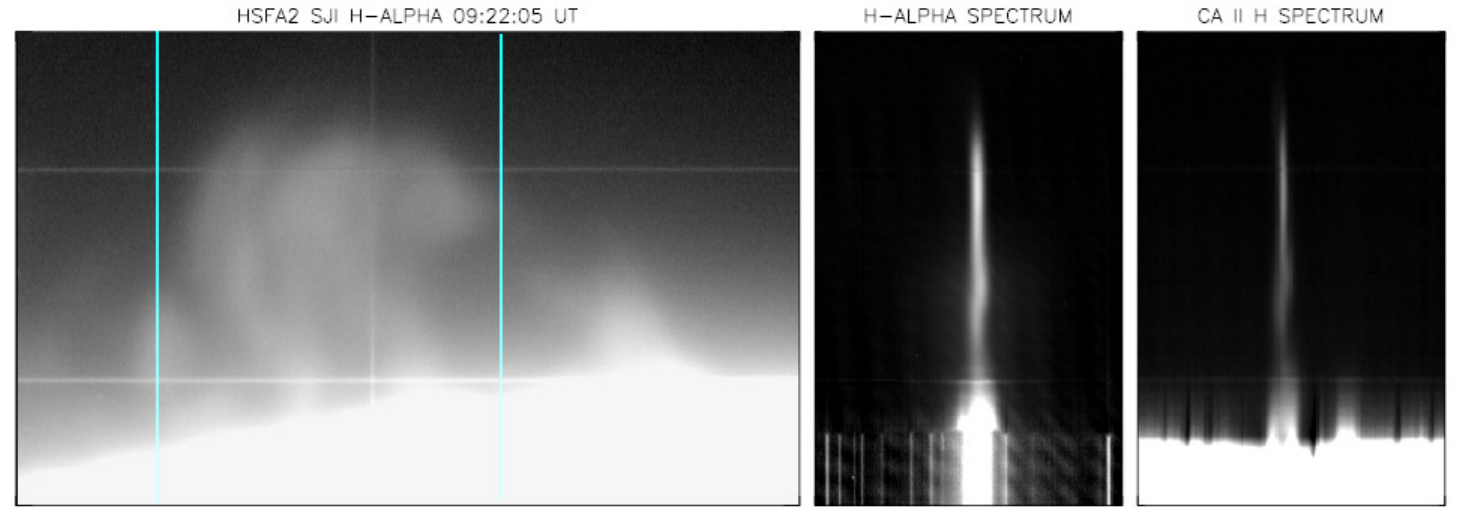}\hfill
  \caption{Example of HSFA2 observations. Left: HSFA2 SJI in H$\alpha$. The vertical line in white corresponds to the position of the spectrograph slit, the two vertical cyan lines mark the two extreme positions of the slit, and two horizontal lines are used for the coalignment of the SJI and spectra. Right: H$\alpha$ and \ion{Ca}{ii} H spectra taken along the vertical slit in the SJI with scattered light removed. SJI and spectra are coaligned in the vertical direction.}
  \label{fig:hsfaiandp}  
\end{figure*}

The quiescent prominence studied in this work was observed during the joint observing campaign HOP 334, conducted from March 27 to April 4, 2017. A long-extended filament began crossing the east limb on March 25; a segment of it was clearly visible and was observed by the international campaign on March 28. This campaign provided coordinated observations from space-borne instruments (SOHO/SUMER, SDO/AIA, and IRIS) and ground-based telescopes (HSFA2 of the Ondřejov Observatory). This campaign was the final opportunity for SUMER observations, as its detector was nearing the end of its operational lifetime. Moreover, the SUMER team agreed to perform challenging observations at the Ly$\alpha$ line, providing in such a way access to the resonance lines of \ion{H}{i}, \ion{Mg}{ii} and \ion{Ca}{ii} by combining SUMER, IRIS, and HSFA2 (the last simultaneous observations of these lines dating back to OSO-8 era \citep{vial82}). Table~\ref{tab:obs} summarizes the observations used in our analysis.

\begin{table*}
\centering
  \caption{Summary of observations of the quiescent prominence on March 28, 2017.}
  \label{tab:obs}
  \begin{tabular}{lllllll}
  \hline\hline
             &   & \multicolumn{2}{c}{Observation time} & & \multicolumn{2}{c}{Radial distance ($R_\sun$)} \\
  Instrument & Data      & Start & End & Cadence & Minimum & Maximum\\ \hline
  IRIS & SJI 1400\,Å, 2796\,Å     &  08:06 & 09:58 & 60s   &0.9 & 1.2 \\
  IRIS & \ion{Mg}{ii}\ spectrum & 09:08 & 09:58 & 30s &0.9 & 1.2\\
  SDO/AIA & 304\,Å       & 09:00 & 12:00 & 12s   &0 & 1.3  \\
  SOHO/SUMER & Ly$\alpha$    & 09:00 & 09:59 & 15s & 0.98 & 1.1 \\
  HSFA2 & SJI H$\alpha$  & 08:54 & 13:38 & 2s &0.95 & 1.1\\
  HSFA2 & H$\alpha$, \ion{Ca}{ii} H spectrum & 08:54 & 13:38 & 2s &0.95 & 1.1 
  \\ \hline
  
  \end{tabular}
\end{table*}
\subsection{IRIS}
IRIS provided high-resolution imaging and spectroscopic observations of this quiescent prominence. The slit-jaw images (SJI) and spectra were taken simultaneously, providing a good opportunity to explore the physical process of solar activities. During the joint observation, IRIS was running very large sparse 32-step rasters with a spatial resolution of $0.33\arcsec$ along the slit, a step size of $2\arcsec$ perpendicular to the slit, and a temporal resolution of $30$s. The field of view (FOV) of the SJI is $167\arcsec\times174\arcsec$. IRIS's observations of this prominence were divided into two segments, starting at 08:06~UT and 09:08~UT, respectively, each lasting 50 minutes. The left panel of Fig.~\ref{fig:irisiandp} shows the IRIS SJI 2796\,Å. The vertical black line in the center corresponds to the position of the spectrograph slit, and the two vertical cyan lines mark the two extreme positions of the raster slit. It is oriented at an angle of about 40 degrees to the main axis of the prominence, and scans the northern portion during the raster observation. At the center of the image, one can clearly observe a cavity surmounted by two strands. The right panel in Fig.~\ref{fig:irisiandp} shows the \ion{Mg}{ii}\ k (2796.4 Å) and h (2803.5 Å) spectra taken along the slit with an exposure time of $30\,\mathrm{s}$. In the current work, we mainly use the SJIs 2796\,Å and the raster data in the doublet of \ion{Mg}{ii} k and h lines. The calibrated level-2 data are used with dark current and flat-field correction, as well as geometric and wavelength calibration \citep{depontieu14}. We performed radiometric calibration using the iris\_get\_response.pro script (version 4) to convert the spectra from data number units to absolute intensity in physical units ($\mathrm{erg\,cm^{-2}\,sr^{-1}\,\AA^{-1}\,s^{-1}}$). The comparison of two different calibration procedures was investigated in \citet{2019AA...624A..72Z}, and a difference of less than 5\% across the whole spectral range was derived. 

\subsection{SUMER}
During the joint observation, the high-resolution SUMER spectrograph recorded a sit-and-stare time series of full Ly$\alpha$ profiles at a cadence of $15\,\mathrm{s}$. The 120 pixels along the slit of 0.3" width were dispersed into 100 spectral pixels recorded on the bare photocathode of the detector spanning a window of $4.3\,\mathrm{\AA}$ around $1215.7\,\mathrm{\AA}$. The observation was preceded by an established sequence needed for exact pointing during non contact hours without real-time telemetry employing onboard hardware encoders. This was to overcome the unreliability of the azimuth drive. Based on the final encoder reading we can assume an uncertainty of the azimuth position of the slit better than 5" which is further corroborated by Fig.~\ref{fig:irisandsumer}. Normally, the slit is pointed interactively and its position monitored by visual inspection.

The slit crossed the target prominence between 08:31~UT and 10:18 UT and was pointed at the east limb in the north-south (NS) direction with about one third of the full length on disk. The slit was not aligned to the NS direction, since the SOHO roll is not readjusted anymore.

Because of the high photon flux, the aperture door had to be partially closed prior to the observation (cf., \citet{curdt2008ly} for details). Thus, the incoming photon flux of the bright line was reduced to a level of about 20\% as accurately determined post factum by two disk exposures in the Lyman continuum around 880\,Å before and after re-opening of the aperture door.

Another limitation was related to the detector. The duty cycle had to be reduced, to avoid overheating of the channel plate. With a duty cycle of 57\%, the entire observation with 69 cycles took 107 minutes.

The observed data were processed with standard SolarSoft procedures \citep{schuhle}. The correction and calibration procedures included decompression, odd even correction, flat field correction, dead time correction, local gain correction and radiometric calibration. The precision of photometric calibration of SUMER has decreased in the last years of observation. We tried to derive the absolute intensity by considering the quiet Sun as a reliable and stable calibration source. We used SUMER Ly$\alpha$ raster observation close to the disk center taken during the joint observing campaign on March 29. The integrated intensity was tentatively compared with the results obtained by \citet{2020Quiet} from the SUMER rasters taken during a minimum of solar activity. They provided Ly$\alpha$ integrated intensities at different solar positions from center to limb. The variation of the solar radiation was taken into account by using the averaged LISIRD Lyman composite index \footnote{\url{https://lasp.colorado.edu/lisird/data/composite_lyman_alpha}} coefficients for the dates of the prominence observation with respect to the date of observations used by \citet{2020Quiet}. 

\subsection{HSFA2}
\begin{figure*}[hbt]
 \centering
  \includegraphics[width=.83\linewidth]{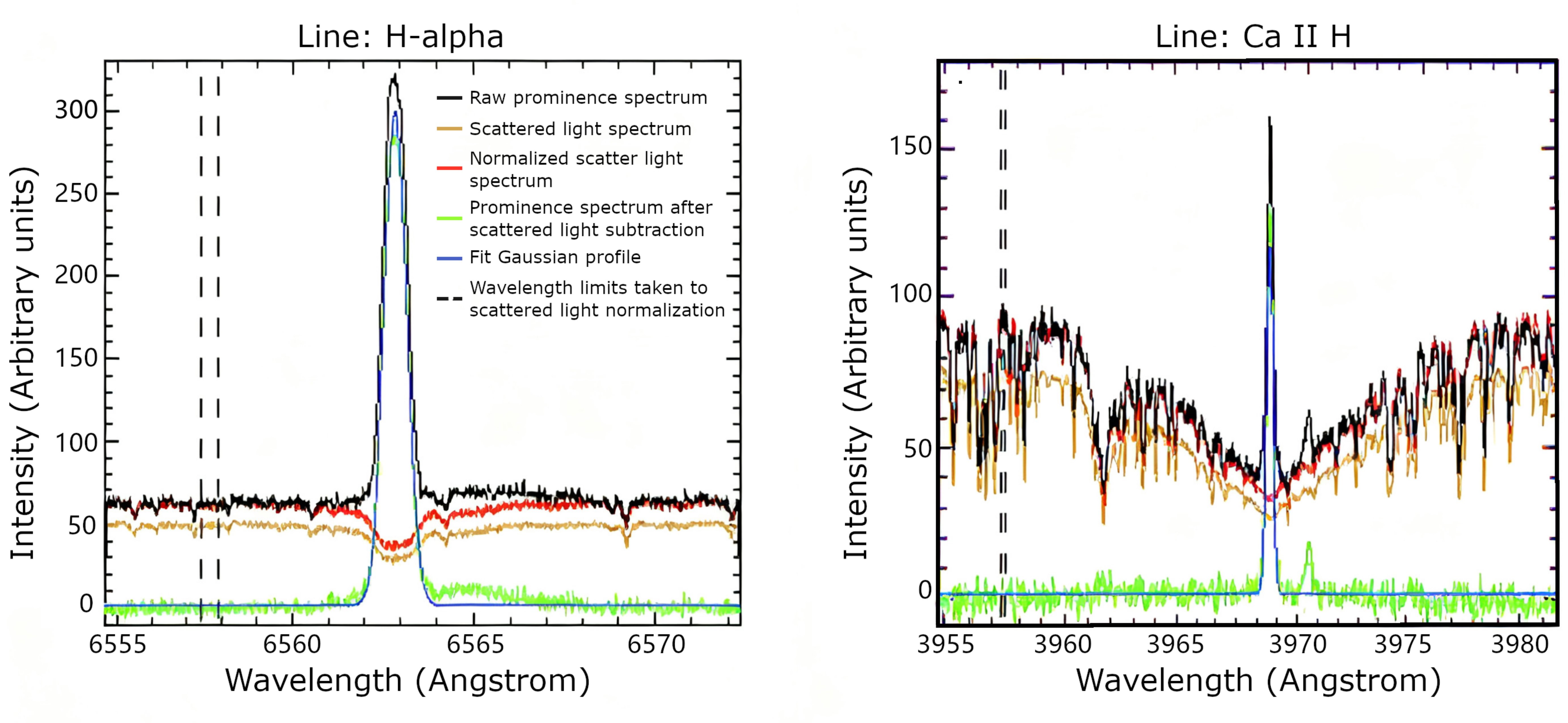}\hfill
  \caption{Example of the scatter light subtraction process for H$\alpha$ and \ion{Ca}{ii} H spectra taken with the HSFA2 spectrograph. Black line: Raw prominence spectrum. Brown line: Scattered light spectrum. Red line: Normalized scatter light spectrum. Green line: Prominence spectrum after scattered light subtraction. Blue line: Fit Gaussian profile. Dashed line: Wavelength limits taken to scattered light normalization.}
  \label{fig:scatt}  
\end{figure*}

The quiescent prominence was also observed by the multichannel solar spectrograph (HSFA2) situated at the Ond{\v r}ejov Observatory, Czech Republic on March 28, 2017 from 08:54 UT to 13:38 UT. The pixel size in the focal plane of the spectrograph is 0.2675'', but the spatial resolution is often limited by seeing to approximately 3''.  The SJIs in H$\alpha$ were taken simultaneously with the spectra in H$\alpha$, H$\beta$ and \ion{Ca}{ii} H. An example SJI of the studied prominence is shown in the left panel of Fig.~\ref{fig:hsfaiandp}. The vertical slit scanned the prominence at 15 positions and the two extreme positions are marked with the vertical cyan lines. Two horizontal lines are used for the coalignment of the slit-jaw and spectral images. 
Exposure time was 120 ms for SJI and 500 ms for spectral images.
The spectra were reduced in a standard way similar to the method described in \citet{2002wohl}. Calibration of observed spectra in wavelength was done by comparing observed positions of photospheric lines on the CCD chip with their reference wavelengths taken from Liège FTS atlas \citep{1973Photometric}. The scattered light samples taken from an area with no structures were carefully subtracted for every spectrum individually. Figure~\ref{fig:scatt} shows the procedure of wavelength calibration and scattered light subtraction.
The middle and right panels of the Fig.~\ref{fig:hsfaiandp}  show the H$\alpha$ and \ion{Ca}{ii} H spectra taken along the slit with scattered light removed. 
Then the absolute intensities in physical units in the prominence were derived by comparing with the absolute intensities in far line-wing of the spectra listed in the table given by \citet{1973allen}.

In the current work, we analyzed the H$\alpha$ and \ion{Ca}{ii} H observations taken between 09:09 UT and 09:13 UT in coordination with the IRIS observations. The H$\beta$ spectra were not used because of the low signal-to-noise ratio. 

\begin{figure*}[hbt]
 \centering
  \includegraphics[width=.83\linewidth]{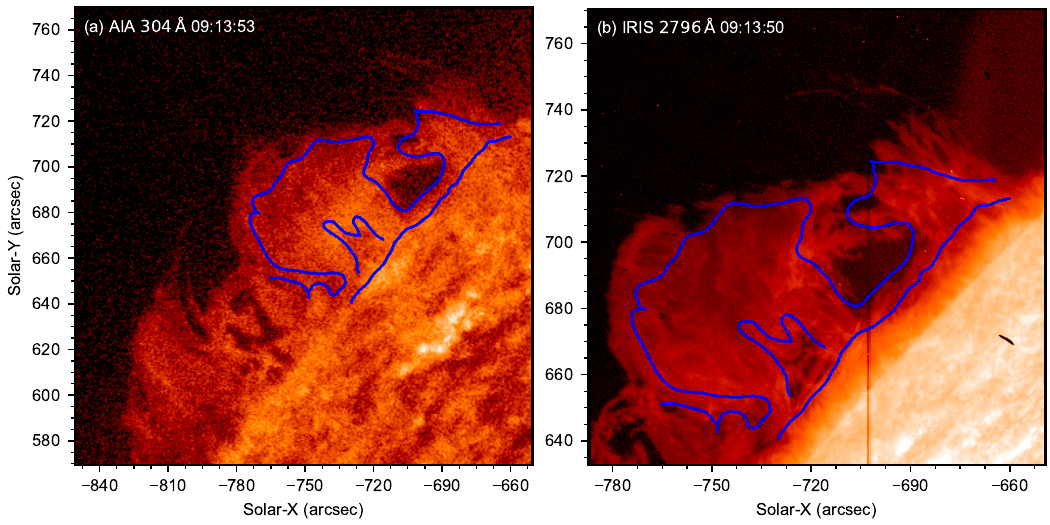}\hfill
  \caption{Images of AIA $304\,\mathrm{\AA}$ and IRIS $2796\,\mathrm{\AA}$ at 09:13\,UT. The inner and outer prominence boundaries and the limb position in H$\alpha$ observed by HSFA2 at 09:22\,UT are marked in blue.}
  \label{fig:irisandhsfap}
\end{figure*}

\subsection{SDO/AIA}
Simultaneous full-disk \ion{He}{ii} 304\,Å observations were made by the SDO/AIA. AIA provides images with a cadence of 12\,s and a spatial resolution of about $1.5\arcsec$. The FOV reaches 1.3 solar radii. The \ion{He}{ii} 304\,Å channel is sensitive to chromosphere and transition region temperatures and was chosen for the identification of the studied prominence in EUV images. The AIA 304 image of this polar crown quiescent prominence was taken at 09:13 UT on March 28, as shown in Fig.~\ref{fig:irisandhsfap}a. 

\subsection{Coalignment of HSFA2 and IRIS images and spectra}
Due to the uncertainty of the pointing precision, the solar X and Y coordinates of HSFA2 may be shifted by up to 30''. The FOV of the HSFA2 is inclined by an angle of -41$^{\circ}$ toward the direction of the IRIS slit. In the current work, the coalignment between IRIS  \ion{Mg}{ii} k and HSFA2 H$\alpha$  SJIs was achieved with the use of AIA \ion{He}{ii} 304\,Å images. The most conspicuous images in H$\alpha$ throughout the joint observations were taken with HSFA2 around 09:22\,UT, as shown in the left panel of Fig.~\ref{fig:hsfaiandp}. We first coaligned the HSFA2 images taken at different times during the observation with the HSFA2 image taken at 09:22\,UT. Then the HSFA2 images and IRIS images were positioned in the larger AIA 304\,Å images by overlapping the bright structures visible. By doing so, we could fit the IRIS SJI with the HSFA2 images in H$\alpha$. Figure~\ref{fig:irisandhsfap} shows the prominence boundaries and disc-limb position in HSFA2 H$\alpha$ overlying the AIA $304\,\mathrm{\AA}$ and IRIS $2796\,\mathrm{\AA}$ images. The prominence in the EUV and UV observations is more extended than the prominence in H$\alpha$. The accuracy of the coalignment is better than 1'' between IRIS and AIA images and is better than 2'' between HSFA2 and AIA images. 

\subsection{Coalignment of SUMER and IRIS images and spectra}
The solar X and Y coordinates of SUMER may be shifted by up to 15''. The alignment between the IRIS and SUMER observations was achieved by comparing the intensity curves. The intensity curve in SUMER observations was computed as the integration of individual spectra over the wavelength in all positions along the slit. By varying the position and the orientation of the SUMER slit in the IRIS SJIs, a large number of curves of intensity along the slit were derived. Based on correlation coefficient between the intensity curves from IRIS and SUMER observations, the inclination angle of SUMER slit was determined to be around 5.3$^{\circ}$, as shown in the left panel in Fig.~\ref{fig:irisandsumer}. The IRIS SJI taken at 09:13\,UT is shown in red, and the dash-dotted line in white represents the position of the SUMER slit. The right panel shows how the integrated intensity changes along the slit. The blue curve corresponds to the IRIS SJI intensity along the white line from the bottom to the top. The black curve corresponds to the Ly$\alpha$ integrated intensity at 09:13:33 UT along the SUMER slit. The intensity of each curve is normalized to arrive at a relative intensity. Note that the maxima of the two curves coincide well.

\begin{figure}
 \centering
 \includegraphics[width=0.99\linewidth]{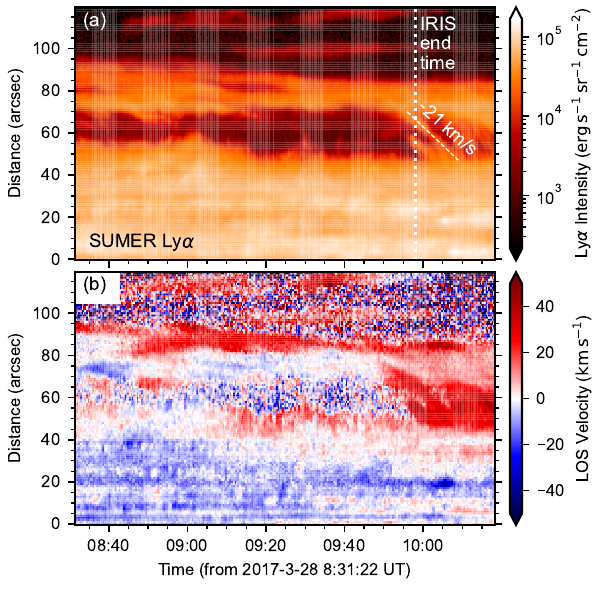}
  \caption{Time-distance diagrams of SUMER Ly$\alpha$ integrated intensity (upper) and LOS velocity (lower). The LOS velocity was derived from Doppler shift using the gravity center method.}
  \label{fig:sumerTD}
\end{figure}

\begin{figure}
 \centering
 \includegraphics[width=0.99\linewidth]{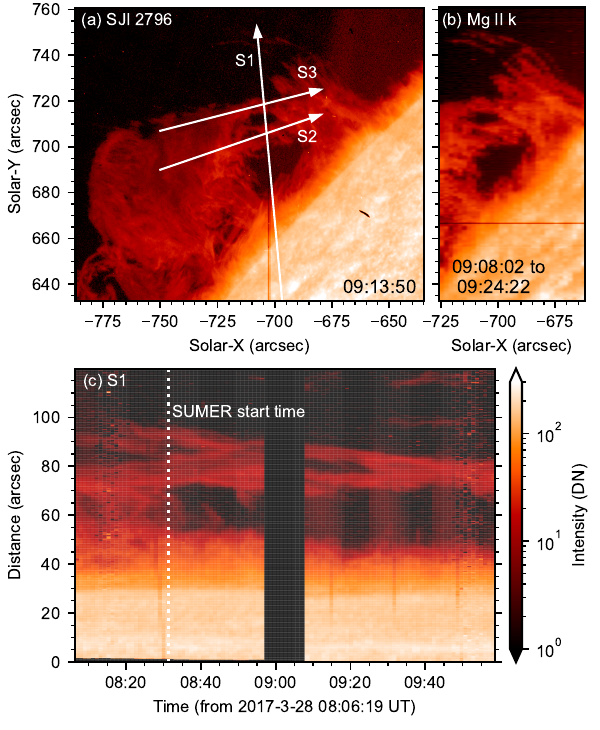}
  \caption{IRIS observations for comparison with Fig.~\ref{fig:sumerTD}. (a) IRIS SJI at $2796\,\mathrm{\AA}$ and the positions of artificial slices for time-distance diagrams. (b) Image of integrated intensity of the \ion{Mg}{ii} k line observed between 09:08:02 and 09:24:22~UT from left to right. (c) Time-distance diagram of IRIS SJI along slice S1, as marked in panel (a). The vertical dotted line represents the starting time of SUMER observations. Note that the traces of raster slit can be seen.}
  \label{fig:irisTD}
\end{figure}

The coalignment of SUMER and IRIS was further validated through their time-distance diagrams. Figure~\ref{fig:sumerTD}a displays the intensity diagram of the SUMER Ly$\alpha$ emission, where the solar disk limb, the cavity between two prominence barbs, and prominence strands above the cavity are clearly visible. The corresponding diagram from the IRIS SJI, along the slice S1 marked in Fig.~\ref{fig:irisTD}a, is shown in Fig.~\ref{fig:irisTD}c. The similar morphological features and dynamic motions of spicule and prominence strands indicate a well-defined coalignment. Notable discrepancies arise from the differing optical thicknesses and spatial resolutions of the two channels: faint prominence threads above the main strands are distinctly identified in the SUMER Ly$\alpha$ image, whereas the SUMER Ly$\alpha$ data exhibit fewer fine-scale details than the IRIS SJI. It is noteworthy that the IRIS SJI $2796\,\mathrm{\AA}$ bandpass has a full width at half maximum (FWHM) of $4\,\mathrm{\AA}$, which exceeds the typical width of the \ion{Mg}{ii} k line ($<1\,\mathrm{\AA}$). The integrated intensity map of the \ion{Mg}{ii} k line from raster observations is presented in Fig.~\ref{fig:irisTD}b, revealing more structural details than the Ly$\alpha$ image but with a less distinct disk limb compared to the IRIS SJI.

\begin{figure}
 \centering
 \includegraphics[width=0.99\linewidth]{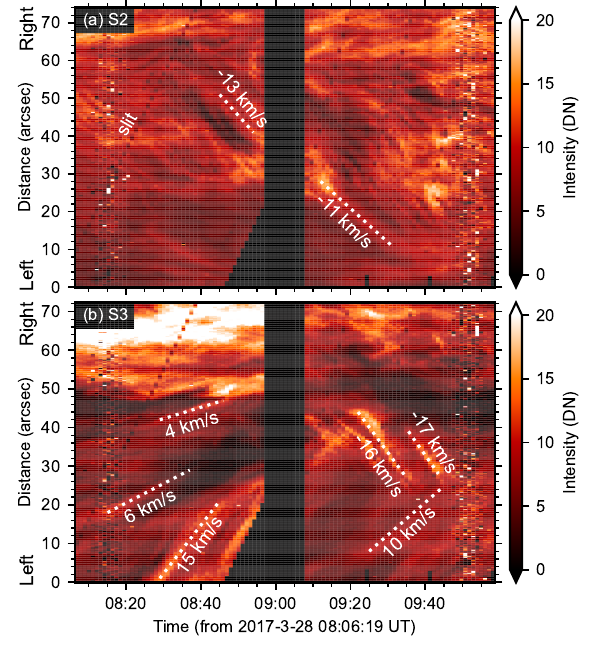}
  \caption{Time-distance diagrams of IRIS SJI $2796\,\mathrm{\AA}$ along slices S2 and S3, as marked in Fig.~\ref{fig:irisTD}a. Some flows are identified with their velocities marked. Note that the spectrograph slit can be seen as dark dotted features along straight lines.}
  \label{fig:flowTD}
\end{figure}
 
\begin{figure}
  \centering
   \includegraphics[width=.99\linewidth]{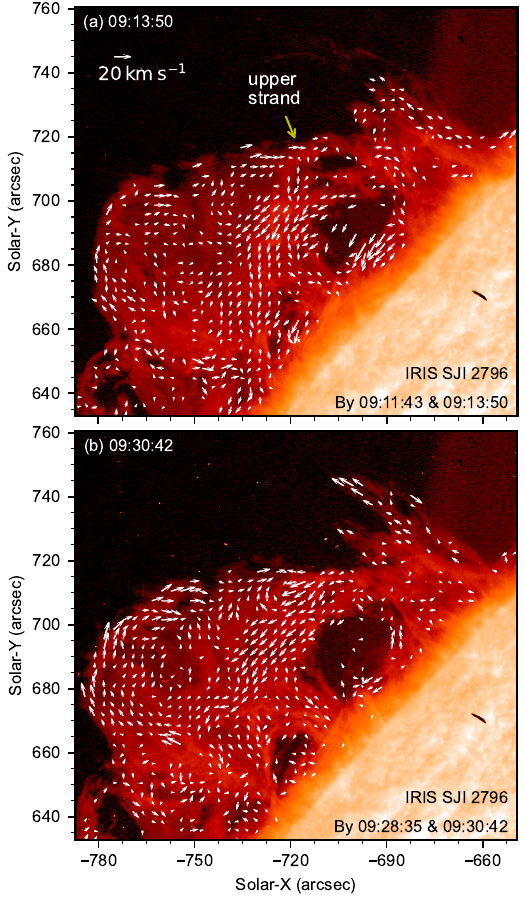}\hfill
   \caption{Plane-of-sky velocities overlying IRIS SJI  $2796\,\mathrm{\AA}$. The velocities are derived from two SJIs using optical flow method, and the observation times are marked at the lower right corner of each panel. The arrow length represents velocity, and an arrow corresponding to a velocity of $20\,\mathrm{km\,s^{-1}}$ is marked in (a).} 
   \label{fig:navevel}  
 \end{figure}
 
\section{Velocities in the prominence}  \label{sec:vel}
The most dynamic segments of the prominence were simultaneously captured by SUMER and IRIS. Using complementary imaging and spectroscopic observations, we derived the three-dimensional velocity fields characterizing the bulk motions of the prominence plasma.

\subsection{Plane-of-sky velocities}

 Plane-of-sky (POS) motions were tracked using both time-slice and optical flow techniques. In the time-distance diagram of Ly$\alpha$ integrated intensity along the SUMER slit, post-IRIS observations revealed downward mass motions from the prominence strands with a characteristic velocity of approximately $-21\,\mathrm{km\,s^{-1}}$. During IRIS observations, the most pronounced flows occurred along the two strands between the prominence barbs. These flows were tracked using artificial slices S2 and S3, as marked in Fig.~\ref{fig:irisTD}a, with corresponding time-distance diagrams from IRIS SJI $2796\,\mathrm{\AA}$ shown in Fig.~\ref{fig:flowTD}. Along the lower strand of slice S2 (Fig.~\ref{fig:flowTD}a), the prominence plasma exhibited dominant right-to-left (negative velocity) flows. Conversely, along the upper strand of slice S3 (Fig.~\ref{fig:flowTD}b), left-to-right (positive velocity) flows were observed before 09:00 UT, which reversed to both opposite flows after 09:08 UT. The change of the flows occurred along with the approach of the two strands. 
 
 The dynamic evolution of the prominence was further characterized using optical flow analysis, as presented in Fig.~\ref{fig:navevel}. Velocity fields were derived via Farneback's algorithm \citep{2003Farneback} implemented in the OpenCV-Python package, leveraging two IRIS SJI observations separated by a $127\,\mathrm{s}$ interval. At 09:13 UT (Fig.~\ref{fig:navevel}a), the lower strand exhibits dominant right-to-left flows with peak velocities of $\sim20\,\mathrm{km\,s^{-1}}$ localized at its left end, while the upper strand displays left-to-right flows. By 09:30 UT (Fig.~\ref{fig:navevel}b), the approaching strands show a reversal in the upper strand's flow direction, now dominated by right-to-left motions. The velocity field of the left part of the prominence exhibits a pattern: downward motions in the lower segment, right-to-left flows in the middle, upward motions along the left edge, and rightward flows in the upper region. This flow topology suggests a mass circulation linking the prominence to the chromosphere and facilitating plasma exchange between adjacent barbs. The observed large-scale, coherent flows challenge the prevailing paradigm of prominence plasma confinement within magnetic dips, implying instead bulk motions between prominence structures.

\begin{figure*}[hbt]
 \centering
  \includegraphics[width=.8\linewidth]{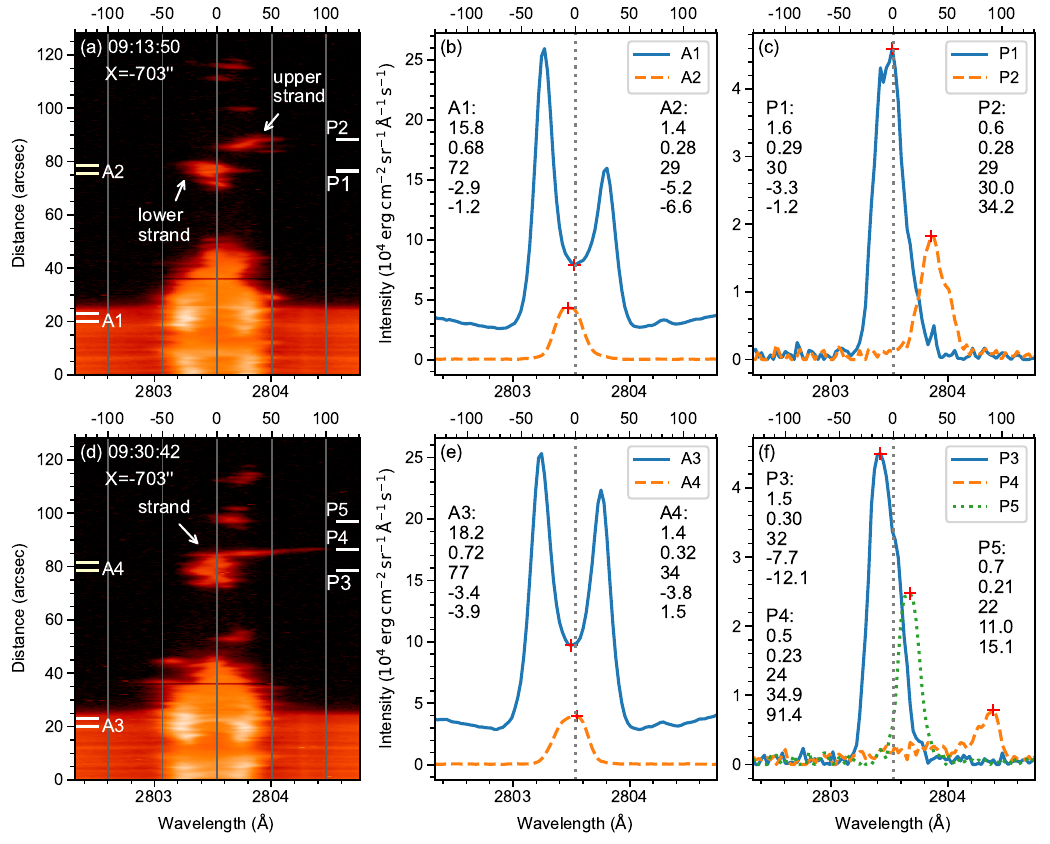}\hfill
  \caption{Spectral observations of the \ion{Mg}{ii} h line. (a) and (d): Stigmatic spectral image. The left bars mark the limits for calculating averaged spectral profiles, and the right bars mark the position of spectral profiles from single spatial pixel. (b) and (e): Averaged profiles. (c) and (f): Profiles from single spatial pixel. The marked profile characters include integrated intensities in $\mathrm{10^4\,erg\,cm^{-2}\,sr^{-1}\,s^{-1}}$, FWHMs in $\mathrm{\AA}$, FWHMs in $\mathrm{km\,s^{-1}}$, LOS velocities by gravity center method in $\mathrm{km\,s^{-1}}$, and LOS velocities by profile centers identified by peaks or valleys in $\mathrm{km\,s^{-1}}$, which are marked with red pluses.}
  \label{fig:mg2hProf}  
\end{figure*}

\subsection{Line-of-sight velocities}

\begin{figure*}[hbt]
 \centering
  \includegraphics[width=.8\linewidth]{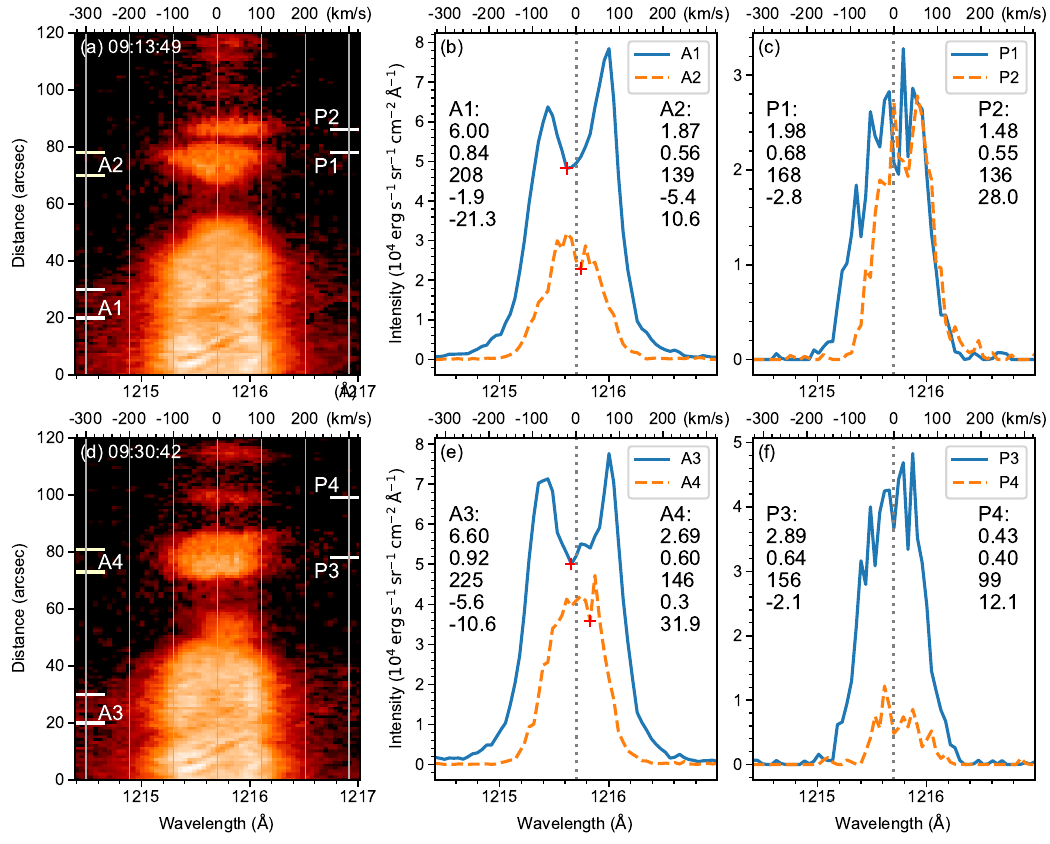}\hfill
  \caption{Similar to Fig.~\ref{fig:mg2hProf} but for SUMER Ly$\alpha$ observations.}
  \label{fig:sumerProf}  
\end{figure*}

\begin{figure*}[hbt]
 \centering
  \includegraphics[width=.8\linewidth]{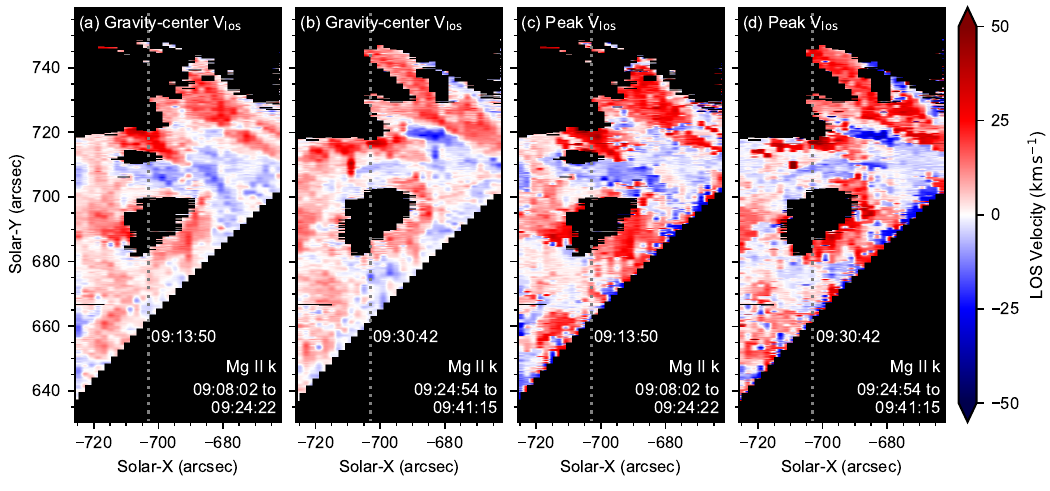}\hfill
  \caption{Doppler images of the \ion{Mg}{ii}\ k line by the gravity center method (a--b) and peak method (c--d)  at two scans. The vertical dotted lines represent the observation times of the \ion{Mg}{ii} h profiles in Fig.~\ref{fig:mg2hProf}.}
  \label{fig:mg2kVlos}  
\end{figure*}
The LOS velocity was derived from spectral line Doppler shifts. Accurately measuring these shifts for \ion{Mg}{ii} h\&k and \ion{H}{i} Ly$\alpha$ lines is non-trivial due to potential optical thickness effects in prominences, which often exhibit reversed line profiles \citep{2014A&A...564A.132H}. For optically thick lines, different segments of the line profile sample plasma parameters from varying depths \citep{leenaarts2013formation}. Even for optically thin lines, multiple components with different velocities along the LOS, showing multi-peak profiles, make the calculation of LOS velocity complicate. In the latter case, gravity center method is generally used to obtain an averaged LOS velocity.

The reference wavelength in the IRIS \ion{Mg}{ii} channel was obtained by averaging disk emission and assumed that the Doppler shifts of the photospheric \ion{Ni}{i} $2799.5\,\mathrm{\AA}$ and \ion{Mn}{i} $2795.6\,\mathrm{\AA}$ lines were negligible. To investigate the spectral characteristics of the \ion{Mg}{ii} h\&k lines and validate the applicability of the gravity center method in our analysis, we present stigmatic spectral images, spatially averaged profiles, and single-pixel profiles in Fig.~\ref{fig:mg2hProf}. The \ion{Mg}{ii} h line was selected for detailed analysis because the primary distinction between \ion{Mg}{ii} h\&k lines lies in their intensities rather than profiles, and the \ion{Mg}{ii} k line on the solar disk is contaminated by \ion{Mn}{i} $2795.6\,\mathrm{\AA}$ and \ion{Fe}{i} $2795.8\,\mathrm{\AA}$ lines. The stigmatic spectral images in Fig.~\ref{fig:mg2hProf} correspond to the IRIS slit positions shown in Fig.~\ref{fig:irisTD}a at two distinct times. The slit locations are also marked in Fig.~\ref{fig:mg2kVlos}. Fig.~\ref{fig:mg2hProf}a captures emissions from two strands: the lower strand exhibits a blueshift (motion toward the observer), while the upper strand is redshifted. In contrast, Fig.~\ref{fig:mg2hProf}b features a single strand with varying Doppler shifts, which indicates bidirectional LOS flows.  Above these strands lie two structures, which are clearly visible in the SUMER Ly$\alpha$ observations (Fig.~\ref{fig:sumerTD}). The spatial locations of the averaged profiles are marked by horizontal bars labeled A1–A4 on the left side of the left-column images, with corresponding profiles plotted in the middle column. Single-pixel profiles, marked by bars P1–-P5 on the right side of the images, are presented in the right column. Each profile is characterized by its integrated intensity, FWHM in units of both angstrom and kilometer per second, Doppler velocity derived via the gravity center method, and Doppler velocity determined from the profile center (defined by the valley for reversed profiles and the peak for normal profiles).

The averaged disk emissions of the \ion{Mg}{ii} h line (A1 and A3) exhibit deep reversal and slight blueshift. In contrast, the averaged \ion{Mg}{ii} h profiles from prominence regions (A2 and A4) are single-peaked. Profile A2 shows a slight blueshift of approximately $-6\,\mathrm{km\,s^{-1}}$, whereas the Doppler shift of A4 is complex due to multi-component contributions. The intensity ratio between prominence and disk emissions is $\sim 0.08$, with a spectral width (FWHM) ratio of $\sim 0.43$. In most cases, single-pixel profiles reveal multi-component structures without reversed profiles, indicating the applicability of the gravity center method for deriving average velocities. However, gravity center method tends to provide a lower velocity than peaks. For instance, profile P2 peaks at $34.2\,\mathrm{km\,s^{-1}}$, with its asymmetric shape implying a component exceeding $34.2\,\mathrm{km\,s^{-1}}$, yet the gravity center method yields $30.0\,\mathrm{km\,s^{-1}}$. Profile P4 peaks at $91.4\,\mathrm{km\,s^{-1}}$, but the gravity center method gives only $34.9\,\mathrm{km\,s^{-1}}$. The comparison of the two methods can also be found in Fig.~\ref{fig:mg2kVlos}. 
While flows in prominences are ubiquitous \citep{2015ASSL..415...31E}, velocities exceeding $90\,\mathrm{km\,s^{-1}}$ in quiescent prominences are rarely reported. \citet{2018ApJ...859..121K} found velocities up to $23\,\mathrm{km\,s^{-1}}$ in prominence barbs. \citet{2013ApJ...775L..32A} observed warm flows (approximately $200\,000\,\mathrm{K}$) in active-region filament threads at up to $98\,\mathrm{km\,s^{-1}}$ in EUV $193\,\mathrm{\AA}$ on the disk, and the authors suggested that the flows are related to the magnetic footpoints. \citet{2011ApJ...731...82H} reported falling knots at $\sim 100\,\mathrm{km\,s^{-1}}$, approaching free-fall speeds \citep[see also the review by][]{2015ASSL..415...79K}. Here, we observe high-speed cool prominence plasma at velocity of $91\,\mathrm{km\,s^{-1}}$ with a projected height of $\sim 30\,\mathrm{Mm}$. The origin of such fast flows is worth further studies to understand the mechanisms of dynamic motion in prominences.

Figure~\ref{fig:sumerProf} is similar to Fig.~\ref{fig:mg2hProf} but for SUMER Ly$\alpha$ spectra. Owing to the low signal-to-noise ratio (SNR) of SUMER data, identifying profile centers in single-pixel spectra, even in spatially averaged prominence profiles, is difficult. Hence we only marked the profile centers for averaged profiles. Though the averaged Ly$\alpha$ disk-limb profiles exhibit reversal, they are less pronounced than \ion{Mg}{ii} h emissions. The Ly$\alpha$ width is significantly broader, with a prominence FWHM of $\sim 140\,\mathrm{km\,s^{-1}}$ compared to $\sim 30\,\mathrm{km\,s^{-1}}$ for the \ion{Mg}{ii} h line. This discrepancy arises from both intrinsic physical broadening and instrumental effects. The instrumental broadening can both widen profiles and reduce reversal depths.  The Ly$\alpha$ intensity ratio of prominence over disk is 0.31 for A2/A1 and 0.41 for A4/A3. The dilution factor at (-700'', 700'') is 0.38. The high ratio of prominence over disk emission indicates that enhanced collision in the prominence due to an increase in density and/or temperature is important for Ly$\alpha$ emission.

The wavelength calibration of the SUMER data is difficult because Ly$\alpha$ line was only observed at the solar limb. We firstly averaged Ly$\alpha$ profiles from the same region as A1 but for all the times. Then we defined the line center by the two points at half maximum. This method can avoid the influence of unequal peak values of Ly$\alpha$ profiles, and the calculated Doppler velocities of the prominence are consistent with those derived from \ion{Mg}{ii} lines. At 09:13~UT (Fig.~\ref{fig:sumerProf}a--c), the lower strand is slightly blue-shifted and the upper strand is red-shifted with a velocity of $\sim 28\,\mathrm{km\,s^{-1}}$. At 09:30~UT, the strand does not show a clear Doppler shift, different from varying Doppler shifts revealed by the \ion{Mg}{ii} h line in Fig.~\ref{fig:mg2hProf}d. It is partly due to the fact that SUMER has a lower spatial resolution than IRIS. The two structures above the strand are red-shifted with a velocity of $\sim 12\,\mathrm{km\,s^{-1}}$, similar to that revealed by the \ion{Mg}{ii} h line.

The gravity center method was applied for all the SUMER Ly$\alpha$ profiles, and the derived time-distance diagram of LOS velocity is shown in Fig.~\ref{fig:sumerTD}b. The diagram reveals that the two prominence strands above the cavity are dominated by opposite flows before their coalescence. It suggests that the flows in the two strands have different origins. The origins of the flows in the two strands may be revealed by the Doppler image of the \ion{Mg}{ii} k line in Fig.~\ref{fig:mg2kVlos}. The lower strand is blue-shifted and the upper one is red-shifted, the same as the results of Ly$\alpha$ observations. What is interesting is that the part of the right prominence barb connecting the lower strand is also blue-shifted, and the part of the left barb connecting the two strands is red-shifted. The same direction of LOS motions may indicate that the flow in the lower strand originates from the right barb, and the flow in the upper strand originates from the left barb. This guess is consistent with the POS motions revealed by the time-distance diagrams in Fig.~\ref{fig:flowTD} and optical flows in Fig.~\ref{fig:navevel}. It suggests a mass transport between the two prominence barbs.
 
\begin{figure}
  \centering
    \includegraphics[width=0.85\linewidth]{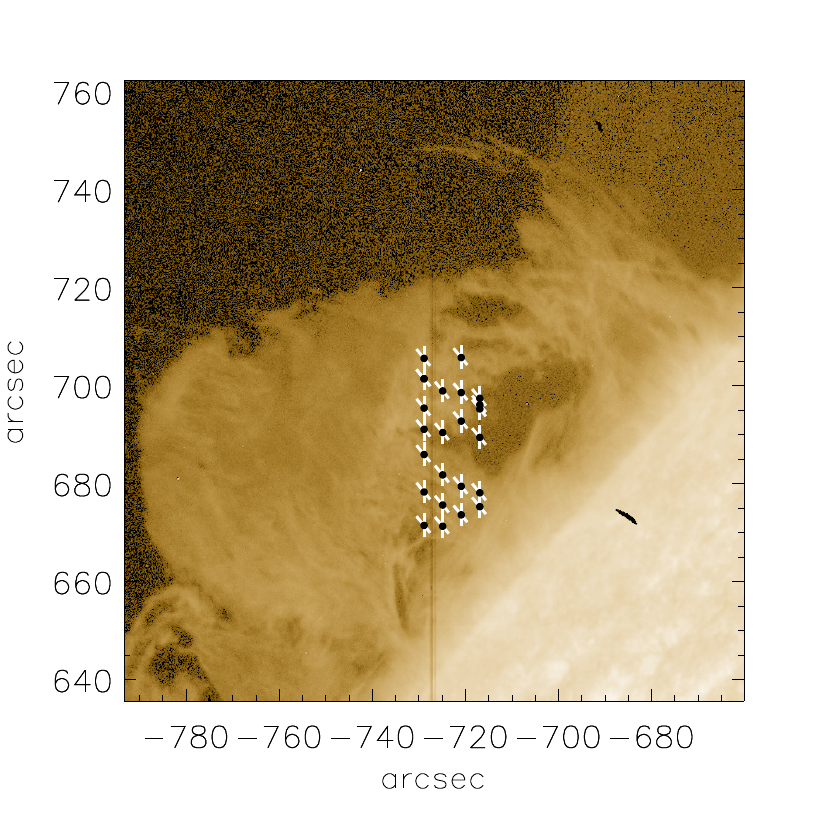}\\[-5mm]
    \caption{Prominence image taken by IRIS at 09:15\,UT, where the black points show the chosen pixels. The short vertical lines in white indicate the pixels within 2.5'' along the IRIS spectrograph slit, and the short inclined lines in white indicate the pixels within 2.5'' along the HSFA2 spectrograph slit.}
    \label{fig:pposition}
   \end{figure}

   \begin{figure}
  \centering
    \includegraphics[width=0.85\linewidth]{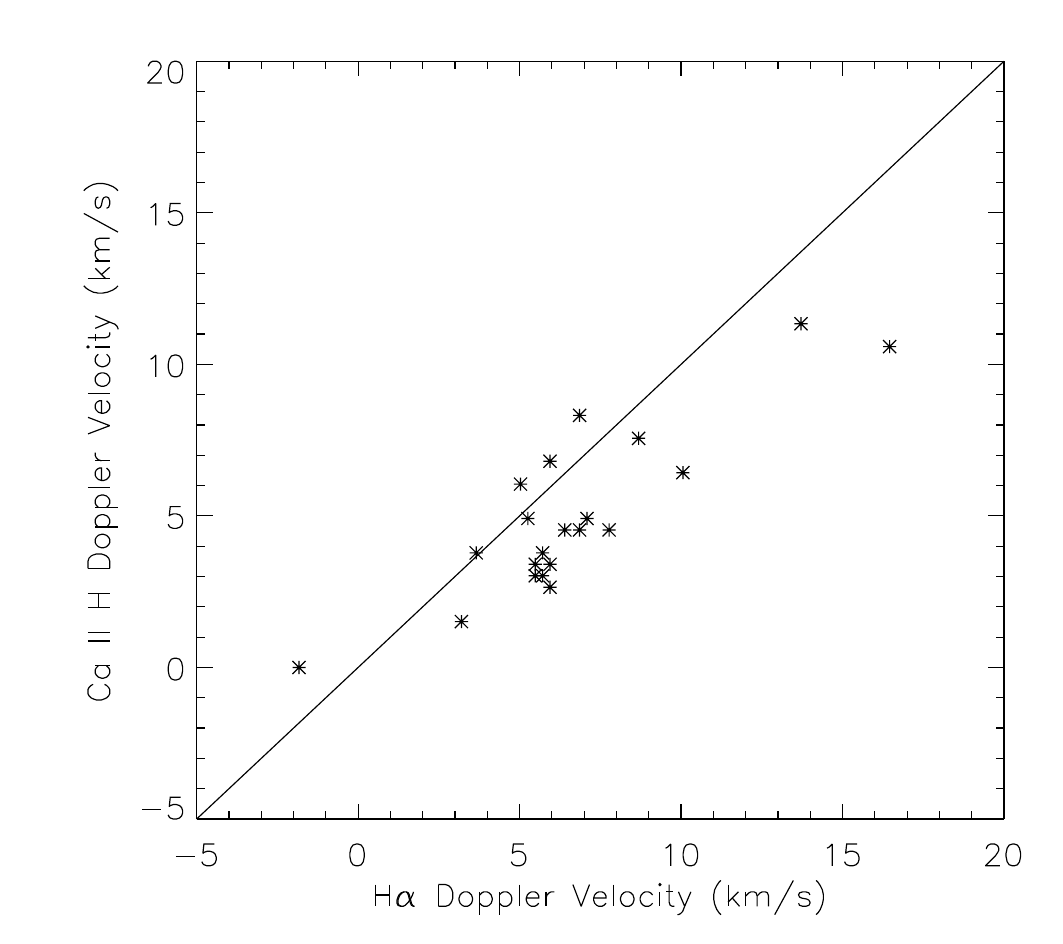}\\[-5mm]
    \includegraphics[width=0.85\linewidth]{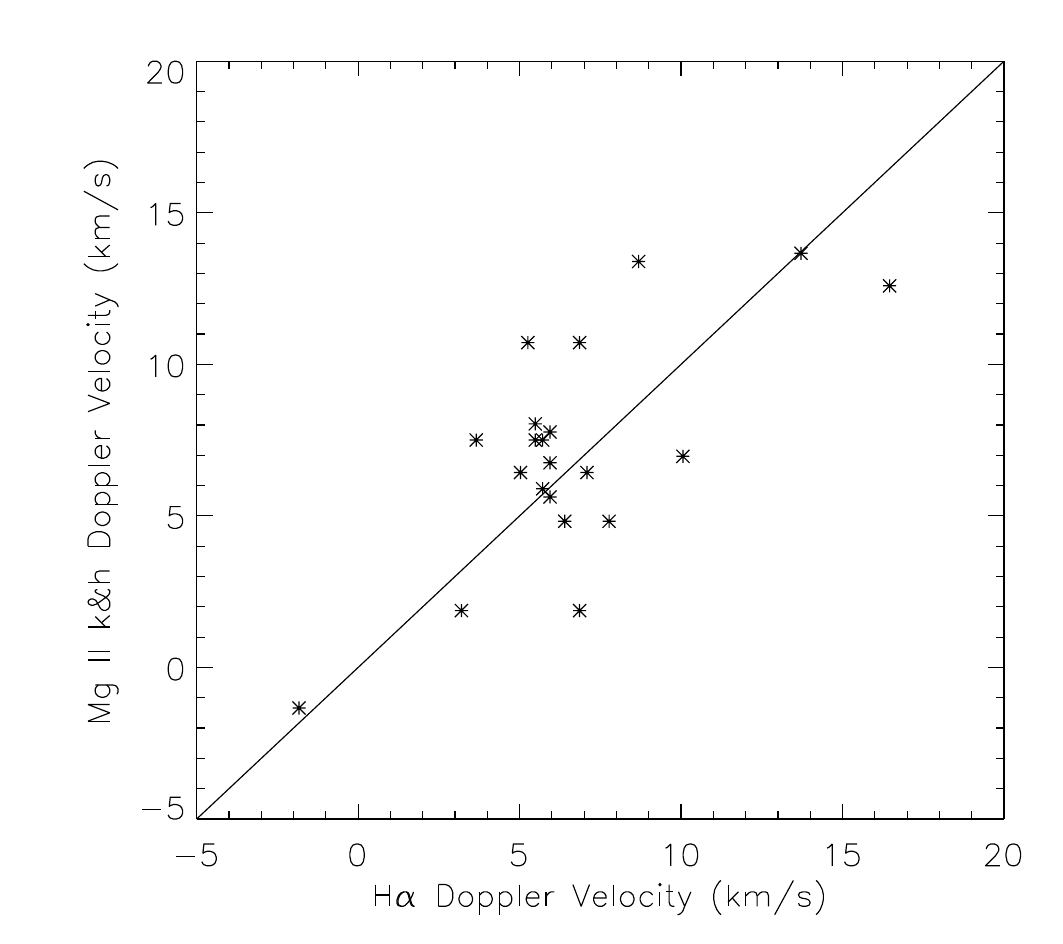}
    \caption{Relation between Doppler shifts derived from 23 chosen points at different wavelengths. }
    \label{fig:euvi22}
  \end{figure}
   
\section{Diagnostic of velocity, density, temperature, and ionization degree in regions jointly studied by space- and ground-based instruments} \label{sec:dia}

\begin{figure*}[hbt]
  \includegraphics[width=.9\linewidth]{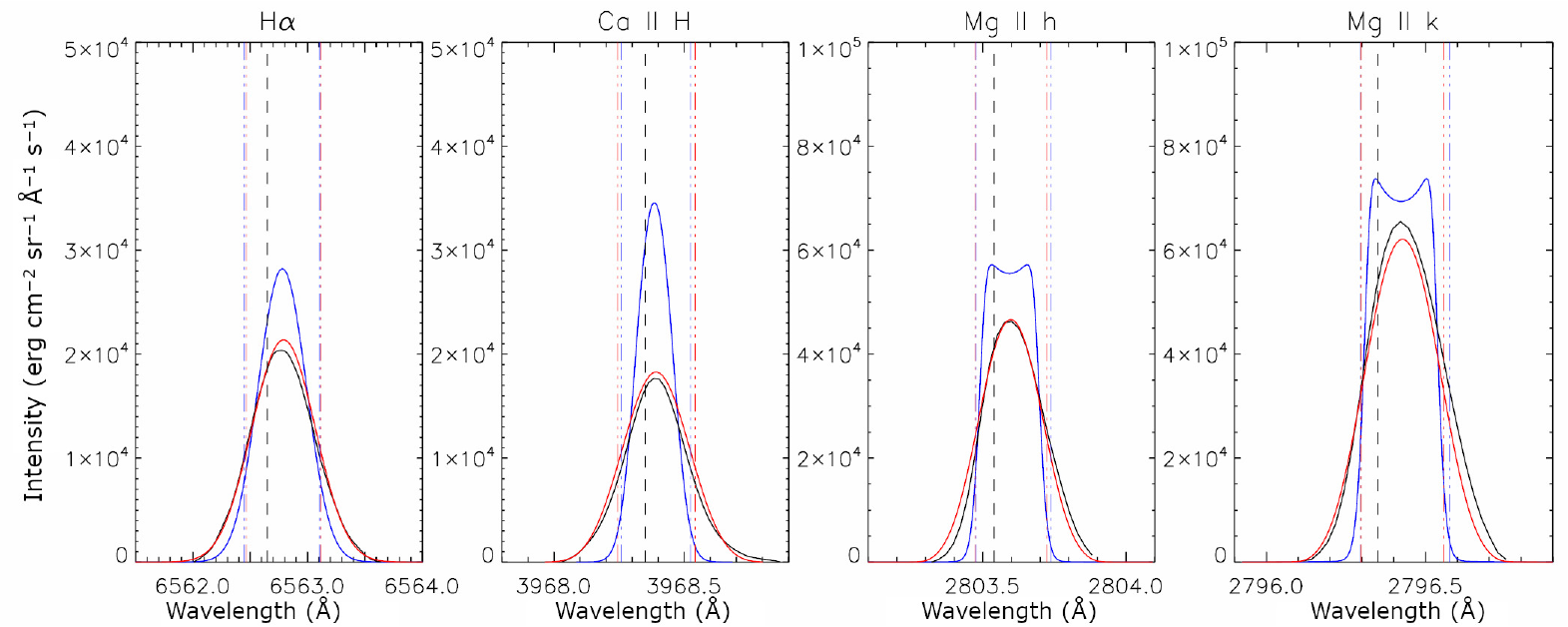}\hfill
 \caption{Comparison of the observed and modeled profiles at (-725", 672"). The black lines show the observed profiles, the blue lines show the modeled profiles, the red lines show the modeled profiles convolved with a macro-velocity of 20\,km\,s$^{-1}$, and the vertical dash-dotted lines show the FWHM positions of each modeled profile. The vertical dashed lines show the position of the line center at rest wavelength.}
  \label{fig:comparep}  
\end{figure*}

\begin{figure}
  \centering
    \includegraphics[width=\linewidth]{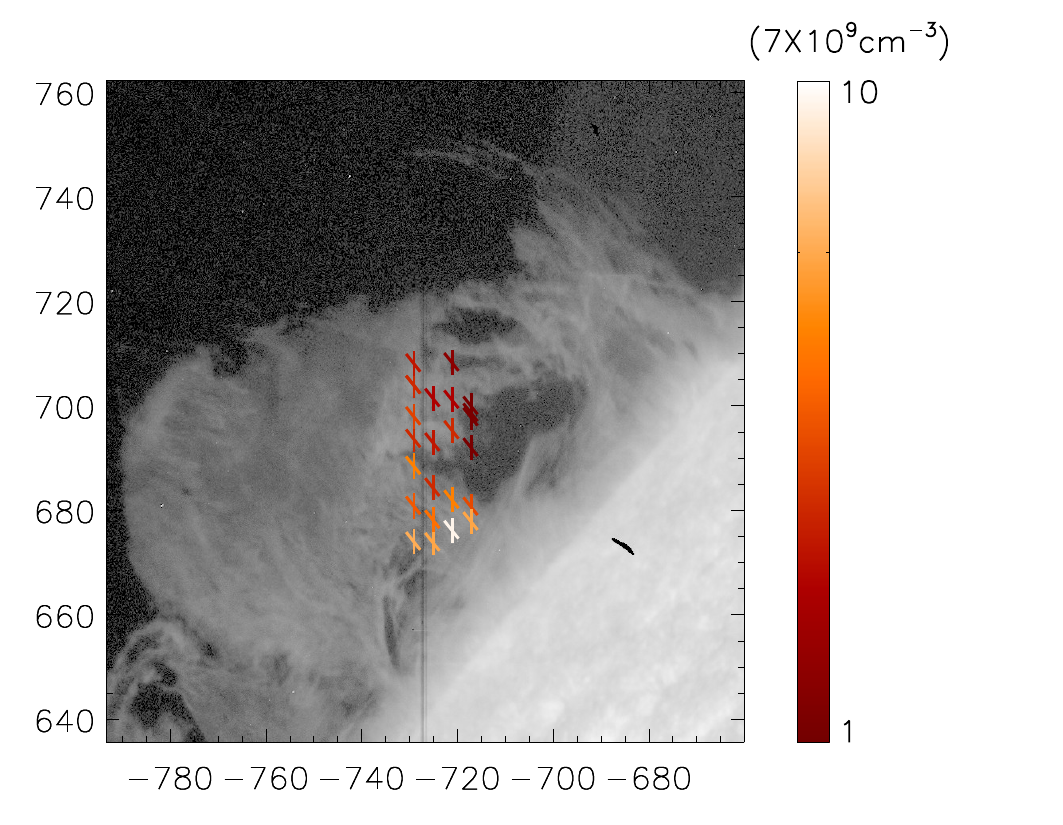}\\[-5mm]
    \includegraphics[width=\linewidth]{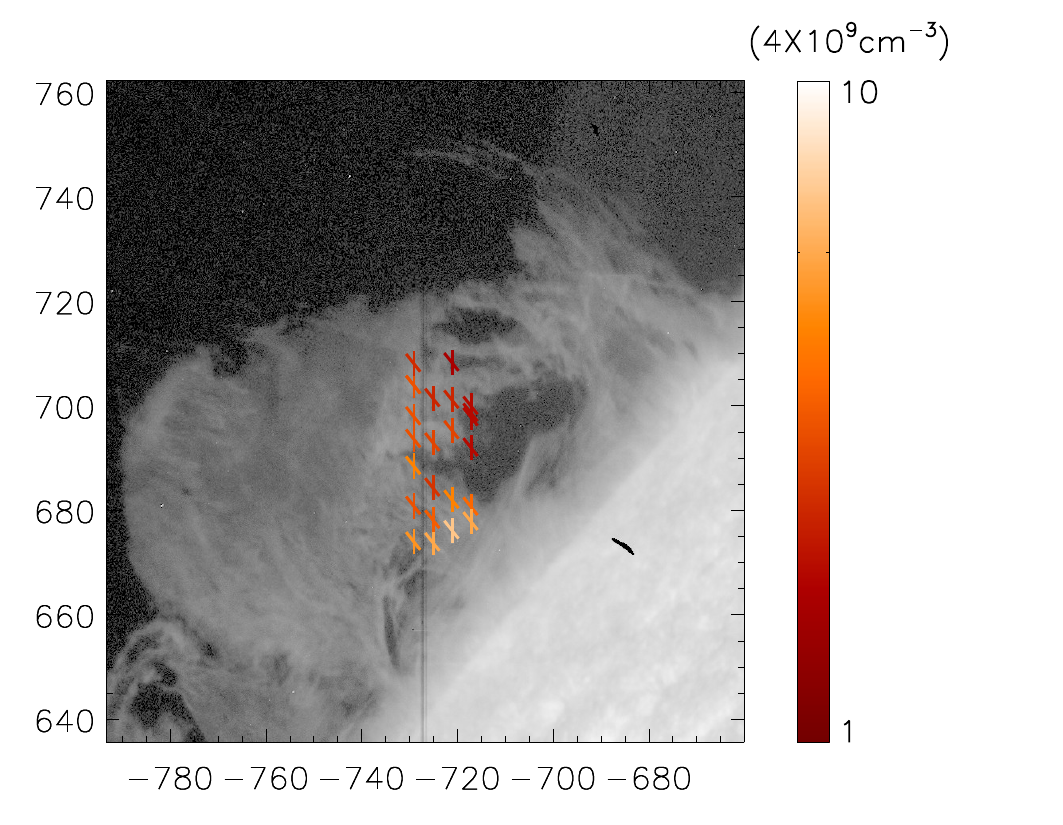}
    \caption{Prominence images taken by IRIS at 09:15\,UT. The colors of the short segments indicate the derived values of hydrogen density (upper panel) and electron density (lower panel).}
    \label{fig:compare1}
\end{figure}
  
 During the prominence observation, the IRIS spectral data were obtained by scanning the region from east to west in 32 steps, while the scan of the HSFA2 proceeded from east to west at 15 positions. Based on the coalignment of different instruments, 23 points where the spectrograph slits of HSFA2 and IRIS intersect in the prominence were derived. We thus obtained a set of 23 H$\alpha$, \ion{Ca}{ii} H, and  \ion{Mg}{ii} profiles in coincident pixels taken simultaneously.  Considering the uncertainty of the co-alignment, for each chosen point, we derive the weighted average profiles of all the pixels within $2.5\arcsec$ from the chosen point along the individual slit. The chosen points are shown in Fig.~\ref{fig:pposition} as black points in the prominence observed by IRIS at 09:15\,UT. The short vertical lines in white indicate all the pixels within $2.5\arcsec$ from the chosen point along the IRIS spectrograph slit, and the short inclined lines indicate all the pixels within $2.5\arcsec$ from the chosen point along the HSFA2 spectrograph slit. The averaging would mix emissions from different threads, and would affect the precision of the derived plasma parameters. However, due to the limited spatial resolution and coalignment accuracy of HSFA2 images, the averaging of profiles is necessary.

 There have been some efforts to combine observations of H$\alpha$ and EUV lines to study the variation of velocities in prominence plasma formed at different temperatures, but they have been somewhat limited by the lower resolution of the EUV data. \citet{cirigliano04a} found an increase in velocity with temperature using Doppler measurements. Figure~\ref{fig:euvi22} shows the relation between Doppler shift derived at the 23 chosen points in different wavelengths. It indicates that spectral lines with rather different formation temperatures are emitted in regions with similar macroshifts \citep[in agreement with ][]{1993A&A...273..267W}. The consistent Doppler velocities derived from H$\alpha$, \ion{Ca}{ii} H and \ion{Mg}{ii} h\&k lines suggest that all of the four lines are emitted in the same position, i.e., core of the prominence. The same emission source is a prerequisite of subsequent spectral inversion. The low LOS velocities, mainly $<15\,\mathrm{km\,s^{-1}}$, indicate that the prominence is quite stable.

The model calculations by \citet{1978ApJ...221..677H} showed that a judicious choice of a restricted number of H, He, and \ion{Ca}{ii} lines may be sufficient for the determination of all relevant atmospheric parameters of the cool prominence matter. We performed plasma diagnostics by comparing H$\alpha$, \ion{Ca}{ii} H,  and \ion{Mg}{ii}\ h\&k FWHM values and integrated intensities from observations and from the NLTE radiative-transfer code PRODOP\footnote{\url{https://idoc.osups.universite-paris-saclay.fr/medoc/tools/radiative-transfer-codes/prodop/}.}.
The prominence structure in this model consists of 1D isothermal and isobaric plasma slabs vertically standing on the solar surface and irradiated by photospheric, chromospheric, and coronal radiation. The input parameters of this code include gas pressure, temperature, the height above the solar surface, geometrical thickness, and the microturbulent velocity. The outputs are physical quantities and optical parameters like electron density, the total hydrogen density, the profiles,  and the optical thicknesses of H, \ion{Ca}{ii}, and \ion{Mg}{ii}\ lines. More details of this code are presented in \citet{2019A&A...624A..56V} . There are no large active regions close to the prominence. Thus we calculated the incident radiation of \ion{Mg}{ii}\ h\&k lines from the absolute intensity of the IRIS full-disk mosaic observation taken on March 28, 2017 for different heights in the prominence. For the incident intensities of other lines, we adopt default settings of PRODOP for simplicity.

To compare with observations, we built a grid of 17820 NLTE models (using the PRODOP code), covering a temperature range of 5000 to 20\,000\,K, a pressure range of 0.005 to 0.5\,dyn\,cm$^{-2}$, and a turbulent velocity range of 5 to 15\,km\,s$^{-1}$. The thickness is set to be 500\,km and 1000\,km given that the cross sections of the small separate threads distributed around (-700", 750") in Fig.~\ref{fig:irisiandp} are about $1\arcsec$-$2\arcsec$. 

\begin{figure*}[hbt]
\centering
  \includegraphics[width=0.99\linewidth]{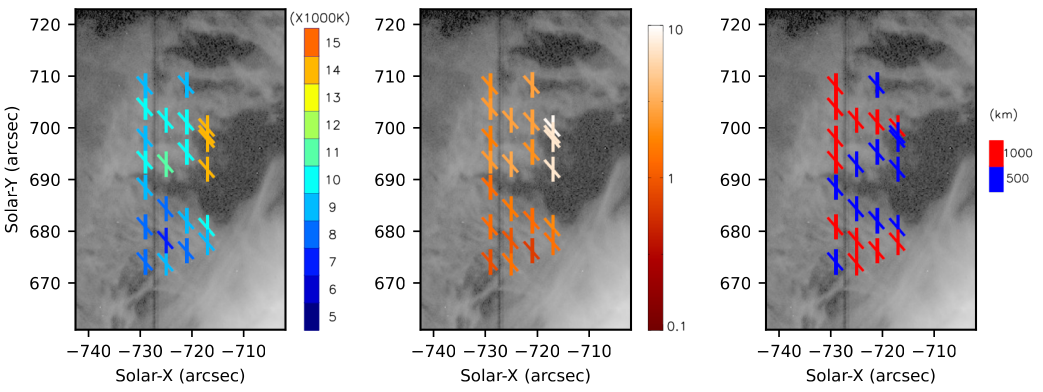}\hfill
  \caption{Distributions of kinetic temperature (left panel), ionization factor as the ratio of electron density to the number density of ground-level hydrogen (middle panel), and geometrical thickness (right panel) of 23 points. The background image is IRIS SJI $2796\,\mathrm{\AA}$ at 09:15\,UT.}
  \label{fig:compare2}
\end{figure*}

With the results of all models, we obtained numerous predicted line profiles of H$\alpha$, \ion{Ca}{ii} H, and \ion{Mg}{ii}\ h\&k lines changing with temperature, pressure, thickness, and turbulent velocity. Before the comparison with observed profiles, the modeled profiles were convolved with Gaussian profiles to decrease the significant inconsistency of widths between the observed and modeled profiles. The differences of the FWHM between the synthetic and observed profiles are computed and the synthetic profiles with a difference of less than 0.1 Å are first chosen. Then we calculated the deviation of line intensities between observed and modeled profiles.
We determined the optimum model for the observations by minimizing the RMS residuals of the four intensities. 
In Fig.~\ref{fig:comparep} we show the comparison between the synthetic and observed profiles for H$\alpha$, \ion{Ca}{ii} H, and \ion{Mg}{ii}\ h\&k lines at (-725", 672"). The black lines show the observed profiles, the blue lines show the modeled profiles, and the red lines show the modeled profiles convolved with a macro-velocity. A better agreement between the synthetic and observed profiles is obtained when the macro-velocity is set to be 20\,km\,s$^{-1}$. The vertical dash-dotted lines show the FWHM positions of each profile. The fit between the  black and red lines looks very promising, while there is a large difference between the black and blue lines. 

Figures~\ref{fig:compare1} and \ref{fig:compare2} show the derived electron densities, total hydrogen densities, and temperatures at the 23 chosen positions in the prominence. The derived electron densities range from $6.5\times10^9$\,cm$^{-3}$ to $2.7\times10^{10}$\,cm$^{-3}$, and the derived total hydrogen densities range from $7.4\times10^9$\,cm$^{-3}$ to $6.6\times10^{10}$\,cm$^{-3}$ in different regions of the studied prominence. The temperature ranges from $7\,000\,\mathrm{K}$ to $14\,000\,\mathrm{K}$. The ionization factor computed as the ratio of the electron density to the number density of ground-level hydrogen is  in the range of 0.68 to 10, which corresponds to the ionization degree (ratio of electron number density to total hydrogen density)  of 0.40 to 0.91. It is understandable that the high ionization values and low densities are obtained at the edges. The densities estimated in previous studies (\citealt{labrosse10b}; \citealt{2019AA...624A..72Z}; \citealt{2014jejcic}) are similar to our values. 

\begin{figure}[hbt]
 \centering
  \includegraphics[width=\linewidth]{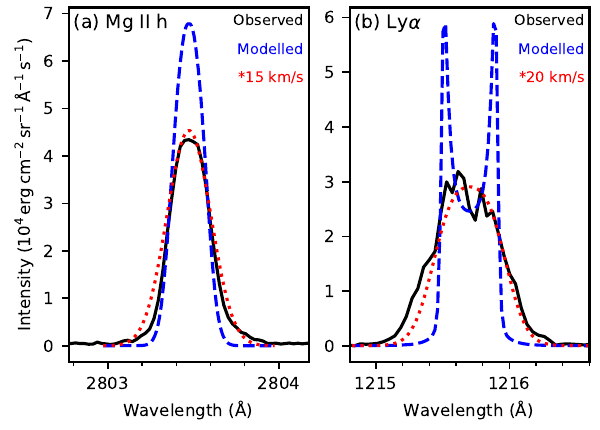}
  \caption{Comparison of the observed and modeled profiles. The black lines show the observed profiles (A2 in Figs.~\ref{fig:mg2hProf} and \ref{fig:sumerProf}), the blue lines show the modeled profiles, the red lines show the modeled profiles convolved with a macro-velocity of $15\,\mathrm{km\,s^{-1}}$ (a) and $20\,\mathrm{km\,s^{-1}}$ (b), respectively.}
  \label{fig:profMgLa}  
\end{figure}

We also tried to fit the average spectral profiles of the \ion{Mg}{ii} h and \ion{H}{i} Ly$\alpha$ lines emergent from the prominence (A2 in Figs.~\ref{fig:mg2hProf} and~\ref{fig:sumerProf}). The observed, synthetic, and convolved profiles are shown in Fig.~\ref{fig:profMgLa}. The height and thickness of the prominence model are set to be $26\,000\,\mathrm{km}$ and $500\,\mathrm{km}$, respectively, and the derived plasma parameters are electron density of $1.0\times10^{10}$\,cm$^{-3}$, total hydrogen density of $1.1\times10^{10}$\,cm$^{-3}$, kinetic temperature of $13\,000\,\mathrm{K}$, and microturbulent velocity of $10\,\mathrm{km\,s^{-1}}$. The modeled \ion{Mg}{ii} h profile is single-peaked, and the Ly$\alpha$ profile is deeply reversed (dashed blue lines). With a Gaussian convolution of $15$ and $20\,\mathrm{km\,s^{-1}}$, respectively, the synthetic profiles (dotted red lines) can well fit the observed ones (solid black lines). These plasma parameters are located within the ranges derived from the above mentioned 23 points. In addition, the optical thicknesses of the \ion{Mg}{ii} h and Ly$\alpha$ lines are 1.2 and 4000, respectively.

\section{Discussion and conclusions}  \label{sec:dis}
We analyzed a fairly complete set of observations of a quiescent prominence on March 28, 2017, including space-based observations from IRIS, SOHO, SDO, and ground-based observations from HSFA2. With the imaging and spectroscopic observations, 3D velocities of bulk motions, relationships between LOS velocities derived from different spectral lines, and thermodynamic parameters of prominence plasma were derived. 

Plane-of-sky velocities derived from both optical flows and time-slice methods revealed mass cycles between the prominence and solar disk, and between the two prominence barbs. Especially opposite flows between the prominence barbs through two strands were observed by both IRIS and SUMER. The upper strand was red-shifted, and mass mainly flowed from left to right; while the lower strand was blue-shifted and dominated by leftward flows. The largest LOS velocity in the strand reached $>90\,\mathrm{km\,s^{-1}}$, which was detected in the \ion{Mg}{ii} h line by IRIS. The Ly$\alpha$ intensity ratio of prominence over disk emissions could reach 0.41, larger than the dilution factor of 0.38, suggesting that plasma collision in prominence contributed to the emission.

We selected 23 points with both IRIS and HSFA2 observations to derive the relationships between LOS velocities of different lines, and to calculate plasma thermodynamic parameters. The H$\alpha$, \ion{Ca}{ii} H, and \ion{Mg}{ii} h\&k lines showed similar LOS velocities, which suggested that these lines were formed at similar depth and were emergent from cool core of the prominence. To derive the plasma parameters, we built a large amount of prominence models using the 1D NLTE radiative transfer code PRODOP. Observed integrated intensities and FWHM values of the H$\alpha$, \ion{Ca}{ii} H, and \ion{Mg}{ii} h\&k lines were used to constrain the models. To well fit the observational spectral profiles, Gaussian convolution with a width of $20\,\mathrm{km\,s^{-1}}$ on the modeled profiles was necessary.
The derived electron densities ranged from $6.5\times10^9\,\mathrm{cm^{-3}}$ to $2.7\times10^{10}\,\mathrm{cm^{-3}}$, and the derived total hydrogen densities ranged from $7.4\times10^9\,\mathrm{cm^{-3}}$ to $6.6\times10^{10}\,\mathrm{cm^{-3}}$ in different regions of the prominence. The mean temperature was around $10\,000\,\mathrm{K}$, and most of hydrogen atoms were ionized, with the ionization degree in the range of $0.40-0.91$.

The results correspond to those of previous studies of quiescent prominences. Using NLTE radiative transfer techniques, \citet{2014jejcic} obtained a similar range of temperature ($6\,000\,\mathrm{K}-15\,000\,\mathrm{K}$), electron density ($5\times 10^9\,\mathrm{cm^{-3}}-10^{11}\,\mathrm{cm^{-3}}$) and a larger range of effective thickness ($200\,\mathrm{km}-15\,000\,\mathrm{km}$) from visible-light emission and quasi-simultaneous HSFA2 spectra of a quiescent prominence detected during the total solar eclipse of August 1, 2008. Comparable results have also been derived from their investigation of the plasma characteristics of a quiescent prominence observed with \ion{Mg}{ii} h\&k lines taken by IRIS and H$\alpha$ spectra observed at Pic du Midi with the MSDP \citep{2022ApJ...932....3J}. They performed a similar spectral inversion technique as we used here, by comparing the integrated intensity of the H$\alpha$ and \ion{Mg}{ii} k lines, the FWHM of both lines, and the ratio of intensities of the \ion{Mg}{ii} k/h lines with the results of 1D NLTE models. This enabled them to obtain values for the prominence temperature ranging from $5\,000\,\mathrm{K}$ to $18\,000\,\mathrm{K}$, the microturbulent velocity peaking around $8\,\mathrm{km\,s^{-1}}$, the electron density of the order of $10^{10}\,\mathrm{cm^{-3}}$, and the peak effective thickness of $500\,\mathrm{km}$.  Using a more elaborate 1D code and including a PCTR, \citet{2024A&A...686A.291P} derived electron density around $10^{10}\,\mathrm{cm^{-3}}$, a temperature ranging from $10\,000\,\mathrm{K}$ to $20\,000\,\mathrm{K}$, and an ionization degree ranging from 0.5 to 1 from the \ion{Mg}{ii} h\&k observations. The higher temperature and larger ionization degree compared to our results are likely to be caused by the disparity in solutions when using spectra of a single ion from 1D prominence modeling to infer the thermodynamic properties.

Let us discuss the meaning of the macro-velocity in the context of convolution, which is necessary for achieving agreement between the observed and modeled profiles. Note that we cannot produce well-fitted profiles through enlarging the microturbulent velocity in the 1D single-slab model, which urge us to think about the origin of the difference between the observed and modeled profiles. Instrumental broadening is one of the candidates. However, the spectral resolution of the IRIS in \ion{Mg}{ii} channel is $53\,\mathrm{m\AA}$ \citep[$\approx 5.7\,\mathrm{km\,s^{-1}}$, ][]{depontieu14}, which is not enough. From the stigmatic spectra of both \ion{Mg}{ii} h and Ly$\alpha$ lines at the position A2 (Figs.~\ref{fig:mg2hProf} and~\ref{fig:sumerProf}), we can see the variations of spectral shifts along the slits, and they suggest dynamic motions along the LOS. Hence, in addition to the instrumental broadening, limited spatial resolution and dynamic motions between multiple threads along the LOS contribute to the broadening of observed profiles. For SUMER observations, due to the large opacity of the Ly$\alpha$ line, the multi-slab effect should be less significant than the \ion{Mg}{ii} h line, and the impacts of instrumental broadening and limited spatial resolution are dominating. The optical thicknesses of H$\alpha$ and \ion{Ca}{ii} H lines are generally much smaller than the Ly$\alpha$ line, and carry multi-thread signals along the LOS. From the SJI in H$\alpha$ (Fig.~\ref{fig:hsfaiandp}), the HSFA2 instrument also suffers from lower spatial resolution than IRIS due to limited seeing. Our work suggests that a Gaussian (or other profiles) convolution is necessary in single-slab models to fit the observed profiles.

We find rather different thermodynamic conditions from a pixel to another one only distant by a few arcsec. This is probably due to the differences between the different threads that are pinpointed, whether their respective internal conditions or their different illuminations. Analysis of the Ly$\alpha$ observations is limited by their relatively low signal-to-noise ratios and resolutions. However, with the end of life of SUMER, the presented spectroscopic observations in Ly$\alpha$ are unique and will not be possible in the immediate future. High-quality spectroscopic observations of Lyman series are expected in the future in order to reveal plasma parameters especially at the bottom of the PCTR.

\begin{acknowledgements}
 This work is supported by the Strategic Priority Research Program of the Chinese Academy of Sciences, Grant No. XDB0560000, NSFC grant No.12233012, National Key R\&D Program of China 2022YFF0503003 (2022YFF0503000), the project of Solar Polar Observatory GJ11020204, NSFC grant No.12203102. J.X. is funded by Basic Research Program of Jiangsu BK20251705. M.Z. is supported by the project RVO:67985815. IRIS is a NASA small explorer mission developed and operated by LMSAL with mission operations executed at NASA Ames Research Center, and major contributions to downlink communications funded by the Norwegian Space Center (NSC, Norway) through an ESA PRODEX contract. The AIA data are provided courtesy of NASA/SDO and the AIA science team. SOHO is a project of international cooperation between ESA and NASA. 

\end{acknowledgements}

\bibliographystyle{aa}
\bibliography{biblio}

\end{document}